# Low-threshold whispering-gallery-mode microlasers fabricated in a Nd: glass substrate by three-dimensional femtosecond laser micromachining

Jintian Lin[1,5], Yingxin Xu[2], Jiangxin Song[1,5], Bin Zeng[1], Fei He[1], Huailiang Xu[3], Koji Sugioka[4], Wei Fang[2,6], Ya Cheng[1,7]

[1] *State Key Laboratory of High Field Laser Physics, Shanghai Institute of Optics and Fine Mechanics, Chinese Academy of Sciences, P.O. Box 800-211, Shanghai 201800, China*

[2] *Department of Optical Engineering, State Key Laboratory of Modern Optical Instrumentation, Zhejiang University, Hangzhou 310027, China*

[3] *State Key Laboratory on Integrated Optoelectronics, College of Electronic Science and Engineering, Jilin University, Changchun 130012, China*

[4] *Laser Technology Laboratory, RIKEN – Advanced Science Institute, Hirosawa 2-1, Wako, Saitama 351-0198, Japan*

[5] *Graduate School of Chinese Academy of Sciences, Beijing 100039, China*

[6]*Email: wfang08@zju.edu.cn*
[7]*Email: ya.cheng@siom.ac.cn*

**Abstract:**

We report on fabrication of whispering-gallery-mode microlasers in a Nd:glass chip by femtosecond laser three-dimensional (3D) micromachining. Main fabrication procedures include the fabrication of freestanding microdisks supported by thin pillars by femtosecond laser ablation of the glass substrate immersed in water, followed by $CO_2$ laser annealing for surface smoothing. Lasing is observed at a pump threshold as low as ~69 μW at room temperature with a continuous-wave laser diode operating at 780nm. This technique allows for fabrication of microcavities of high quality factors in various dielectric materials such as glasses and crystals.

Optical microcavities that have been extensively used to strongly confine light both spatially and temporally are excellent candidates for a wide range of applications including low-threshold lasing [1,2], nonlinear optics [3], biosensing [4], quantum electrodynamics [5] et al, as summarized in the reviews [6,7]. Among them, silica microtoroid cavities [8] fabricated by two-dimensional (2D) planar lithographic approaches have received significant attention since they can provide extremely high quality (Q) factors and small volume via confining the light to narrow equatorial ring by total internal reflection. Fabrication of high-Q microcavities on a variety of transparent materials not limited to silica glasses or silicon substrates would benefit all the applications mentioned above by fully exploiting the properties of different materials. Recently, femtosecond laser direct writing has been proved as a promising solution for fabricating high-Q microcavities in polymers [9-11] and fused silica [12,13]. In particular, fabrication of microcavities of 3D geometries in a fused silica substrate was achieved using femtosecond laser assisted chemical wet etching approach [12]. In such an approach, the fused silica glass is first scanned by a tightly focused femtosecond laser beam for selectively modifying a predesigned area, and then glass material in the modified area is preferentially removed by wet chemical etching [14]. Unfortunately, this technique can only be used for fabrication of 3D structures in very few glass materials (e. g., fused silica and photosensitive glass) as these glasses can be modified by femtosecond laser irradiation for selectively promoting the etch rate inside the glasses. To overcome this difficulty, in this letter, we demonstrate the fabrication of microcavities in rare earth ions doped glass using water-assisted femtosecond laser micromachining [15]. Since this technique does not require chemical etching for material removal, in principle, arbitrary transparent materials including non-silica glasses and nonlinear glasses and crystals can be used as the substrate materials for microcavity construction. As a first attempt to demonstrate such a capability, we fabricate microcavity lasers in a Nd:glass chip and demonstrate lasing action at a pump threshold as low as 69 μW via continue-wave (CW) excitation.

For fabricating whispering-gallery-mode (WGM) microlasers, a Nd:glass (~1.2 wt. % $Nd_2O_3$, 50~60 mol% $SiO_2$) substrate with a thickness of 1mm was used. Both the upper and bottom surfaces of the glass were polished. The procedure of fabrication consists of (1) femtosecond laser ablation of the glass material which is immersed in water and (2) $CO_2$ laser reflow, as illustrated in Fig.1. During the femtosecond laser ablation, the material around the focal spot was selectively removed, enabling formation of arbitrary 3D structures (e. g., the microdisk supported by a thin pillar) by scanning the focal spot in predesigned region. The femtosecond laser micromachining system consisted of a Ti: sapphire femtosecond laser source (Coherent, Inc., center wavelength: ~800nm, pulse width: ~40 fs, repetition rate: 1 kHz), a computer-control XYZ translation stage with 1-μm resolution, and a focusing objective with a numerical aperture (NA) of 0.85. A circular aperture with a diameter of 5 mm was placed in front of the objective to improve the beam quality, and the power was carefully adjusted with a variable neutral density filter. The beam was focused into the glass sample, which was fixed on the translation stage. The ablation front always contacted with deionized water in order to remove the ablated debris. A layer-by-layer annular ablation from bottom to top with 2.9-μm interval between the adjacent layers was adopted, which have ensured sufficient overlap between the adjacent ablation layers. The laser power was chosen to be ~2.8 mW. After the femtosecond laser ablation, a microdisk with a diameter of 74 μm and a thickness of 12 μm, which was supported by a thin pillar with a diameter of 16.7 μm and a height of 51.3 μm, was formed. With our femtosecond laser operated at 1 kHz, it took ~ 120 min to fabricate such a microdisk by ablation; however this time could be shortened using a femtosecond laser operated at higher repetition rate.

For high-Q microcavties, smooth surfaces are important to minimize the scattering loss. Usually, the average roughness of a femtosecond laser ablated surface is on the order of a few tens of nanometer, which is too big for high-Q microcavity applications. Therefore, a $CO_2$ laser reflow step was adopted by which the surface tension of the melted glass created a smooth surface. After the reflow, the microdisk could reshape itself either to a microtoroid (Fig. 2(a)), or to a spheroidal shape (Fig. 2(b)), depending on the length of heating time. The scanning electron microscope (SEM) images of a spheroid microcavity are shown in Figs. 2(c) and (d), whose diameter was measured as 60.0 μm.

To observe the lasing action, the spheroid microcavity fabricated in Nd:glass was optically excited by a CW laser diode operating at 780 nm. The pump laser was first coupled into a bare fiber (Corning, SMF-28) with one of its ends placed close to the microcavity. When the pump laser exited from the fiber, it propagated in free space with a divergence angle. At the spheroid microcavity, the diameter of the pump beam increased to ~80 μm due to a significant distance between the output end of fiber and the microcavity, which ensured a good overlap between the excitation beam and the microcavity. To measure the emission spectra from the microcavity, a 10 X objective lens with a NA of 0.25 was used, and the lateral image of the microcavity was projected onto the entrance slit of a spectrometer (Andor, model: Du920) with a magnification of 35, as shown in the upper inset of Fig.3(b). A 2D charge coupled device (CCD) array detector mounted at the exit port of the spectrometer could capture a 2D spatial-spectral image of the microcavity emission. Before the spectrometer, an optical filter was used to block the 780 nm pump light. The laser emission images were recorded by CCD when the slit was widely open, and the grating was rotated to the position of the zeroth order diffraction (i.e., reflection). Two bright spots located in the both ends of the equatorial ring were observed, as indicated by arrows in the lower inset of Fig. 3(b). This indicates the nature of WGM emission, since light stored in the cavity could only escape tangentially from the edge of the spheroid. To measure emission spectra, we rotated the grating to the position of the first-order diffraction. And the width of entrance slit was reduced to 90 μm, so that only the emission from a narrow vertical strip region of the microcavity could enter the spectrometer. Accordingly, as we adjusted the horizontal position of the sample, we could only observe significant signals when the left or the right side of the microcavity image overlapped with the slit. The emission spectra collected from one side of the microcavity are presented in Fig. 3(a). At low excitation power, only a broad-spectrum photoluminescence (PL) emission was detected. As the pump intensity increased, discrete peaks started to emerge in the recorded spectra. The intensity of these periodic peaks increased dramatically when pump laser intensity was above threshold. The lasing spectrum in Fig. 3(a) appears to be highly multimode with an azimuthal free spectral range (FSR) of ~3.56 nm, which agrees well with the expected value of 3.78 nm estimated by $\Delta\lambda = \lambda^2 / (n \times C)$ based on fundamental WGMs, where $\lambda \sim 1057.78$ nm is the emission wavelength, $n = 1.57$ is the effective index of microcavity, and $C = 188.5$ μm is the circumference of the microcavity. When the pump power reached 0.55 mW, the full width half maximum of the mode at ~1058.64 nm is 0.34 nm. Figure 3(b) shows the measured output laser power as a function of pump laser energy, and the lasing action occurred when the applied total pump power was 69 μW. Considering the mismatch between the pump laser wavelength (780 nm) and absorption peak of the gain medium (808 nm), as well as the low absorption efficiency of the divergent pump beam, actual threshold could be much smaller.

In conclusion, we demonstrate the fabrication of whispering-gallery-mode microlasers in a Nd:glass chip by femtosecond laser 3D micromachining. The threshold of lasing was measured as low as ~69 μW. Further threshold reduction is possible by evanescent-wave pump at near ~808 nm at cryogenic temperatures. Most importantly, our technique offers potential for fabricating high-Q microcavities in a variety of non-silica glasses (e. g., rare earth ions doped glasses and nonlinear functional materials) with 3D geometries (e. g., microtoroid and spheroid microcavity) in a one-continuous process, which can benefit a broad spectrum of applications ranging from nonlinear optics to quantum information processing.

We thank Dr. Jinping Yao, Lingling Qiao, Wei Chu of Shanghai Institute of Optics and Fine Mechanics, Chinese Academy of Sciences, for the help of collecting spectra and data analysis. The work is supported by National Basic Research Program of China (No. 2011CB808100), and NSFC (Nos. 61275205, 11104245, 11134010, 61108015, 61008011).

**References:**

1. Y.-F. Xiao, C.-H. Dong, C.-L. Zou, Z.-F. Han, L. Yang, and G.-C. Guo, Opt. Lett. **34**, 509 (2009)

2. L. He, Ş. K. Özdemir, and L. Yang, Laser Photonics Rev. **7**, 60 (2013)

3. V. S. Ilchenko, A. A. Savchenkov, A. B. Matsko, and L. Maleki, Phys. Rev. Lett. **92**, 043903 (2004)

4. E. Chow, A. Grot. L. W. Mirkarimi, M. Sigalas, and G. Girolami, Opt. Lett. **29**, 1093 (2004)

5. T. Yoshie, A. Scherer, J. Hendrickson, G. Khitrova, H. M. Gibbs, G. Rupper, C. Ell, O. B. Shchekin, and D. G. Deppe, Nature 432, 200 (2004)

6. K. J. Vahala, Nature **424**, 839 (2003)

7. V. S. Ilchenko and A. B. Matsko, IEEE J. Sel. Top. Quantum Electron. **12**, 15 (2006)

8. D. K. Armani, T. I. Kippenberg, S. M. Spillane, and K. J. Vahala, Nature **421**, 925 (2003)

9. Z.-P. Liu, Y. Li, Y.-F. Xiao, B.-B. Li, X.-F. Jiang, Y. Qin, X.-B. Feng, H. Yang, and Q. Gong, Appl. Phys. Lett. **97**, 211105 (2010)

10. J. F. Ku, Q. D. Chen, R. Zhang, and H. B. Sun, Opt. Lett. **36**, 2871 (2011)

11. T. Grossmann, S. Schleede, M. Hauser, T. Beck, M. Thiel, G. Freymann, T. Mappes, and H. Kalt, Opt. Express **19**, 11451 (2011)

12. J. Lin, S. Yu, Y. Ma, W. Fang, F. He, L. Qiao, L. Tong, Y. Cheng, and Z. Xu, Opt. Express **20**, 10212 (2012)

13. K. Tada, G. A. Cohoon, K. Kieu, M. Mansuripur, and R. A. Norwood, IEEE Photo. Tech. Lett. **25**, 430 (2013)

14. K. Sugioka and Y. Cheng, Lab Chip **12**, 3576 (2012)

15. Y. Li, K. Itoh, W. Watanabe, K. Yamada, D. Kuroda, J. Nishii, and Y. Jiang. Opt. Lett. **26**, 1912 (2001)

**Figure captions:**

Fig. 1 (Color online) Procedures of fabrication of a Nd:glass microcavity by femtosecond laser ablation assisted with water, followed by $CO_2$ laser reflow.

Fig. 2 (Color online) Optical microscope images of (a) microtoroid and (b) spheroid fabricated by femtosecond laser ablation after $CO_2$ laser heating. Insets in (a) and (b): top view of the microcavities, (c) and (d) SEM images of the spheroid microcavity at different viewing angles.

Fig. 3 (Color online) (a) Evolution of the microcavity PL emission spectrum with increasing pump power. (b) Laser output power as a function of the pump power, showing a lasing threshold of 69 μW. Upper inset in (b): schematic of the laser experimental setup. Lower inset in (b): laser spots on the edges of microcavity captured with the 2D CCD array detector mounted on the spectrometer.

Fig. 1

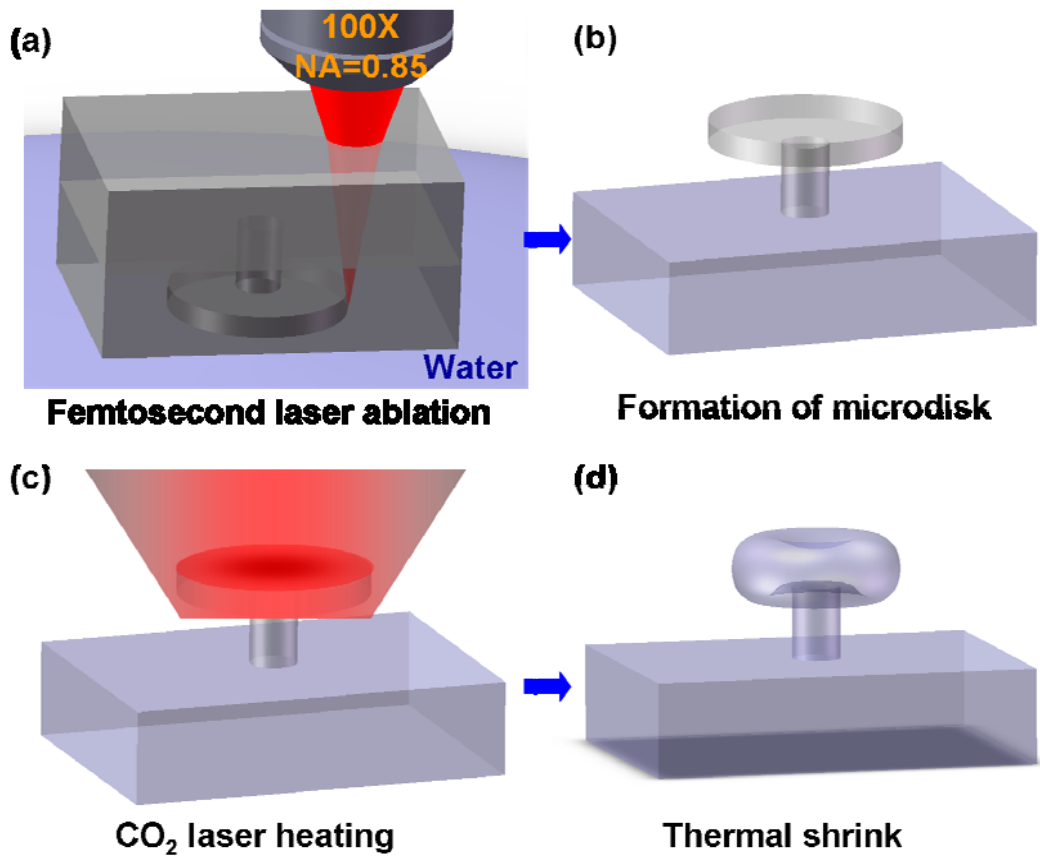

Fig. 2

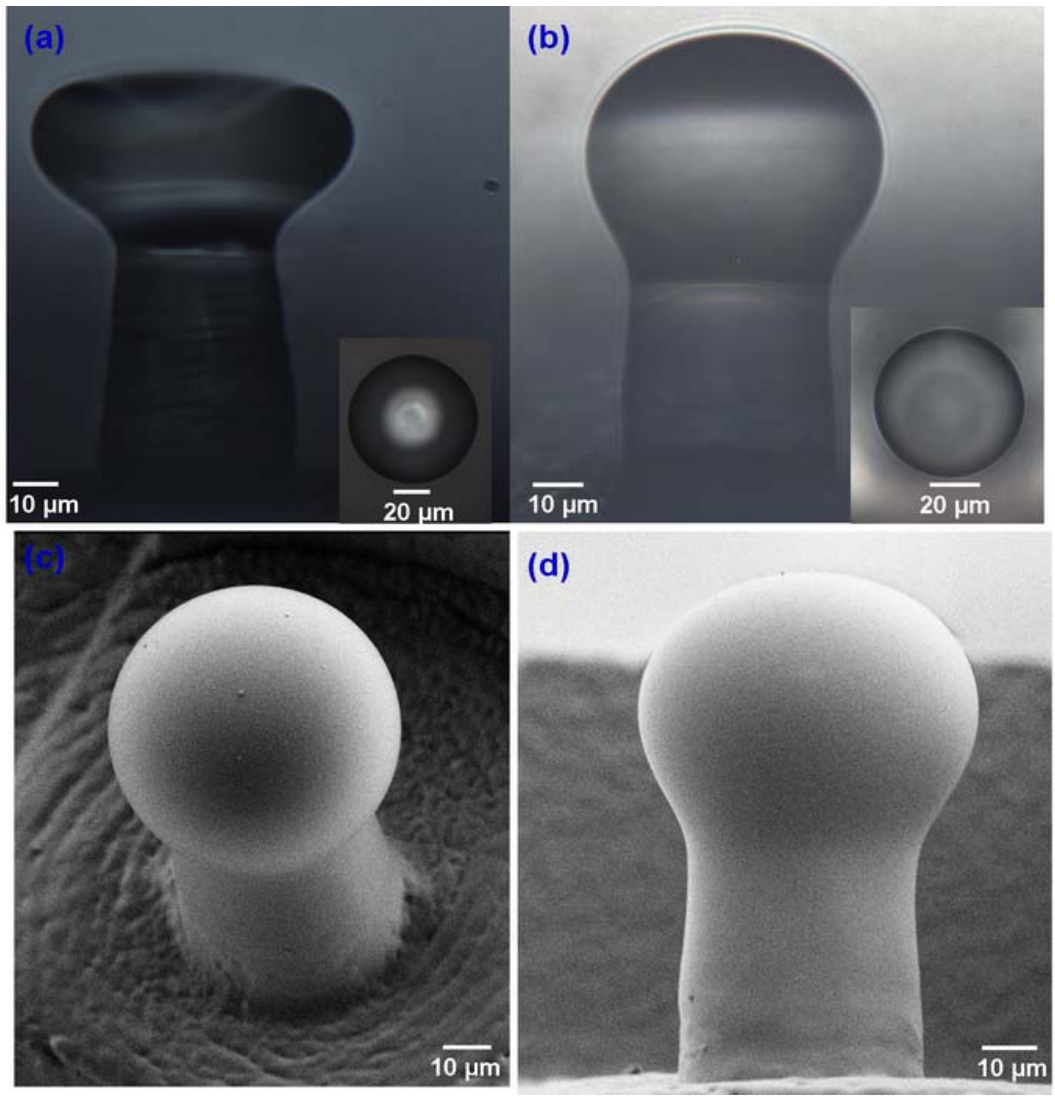

Fig. 3

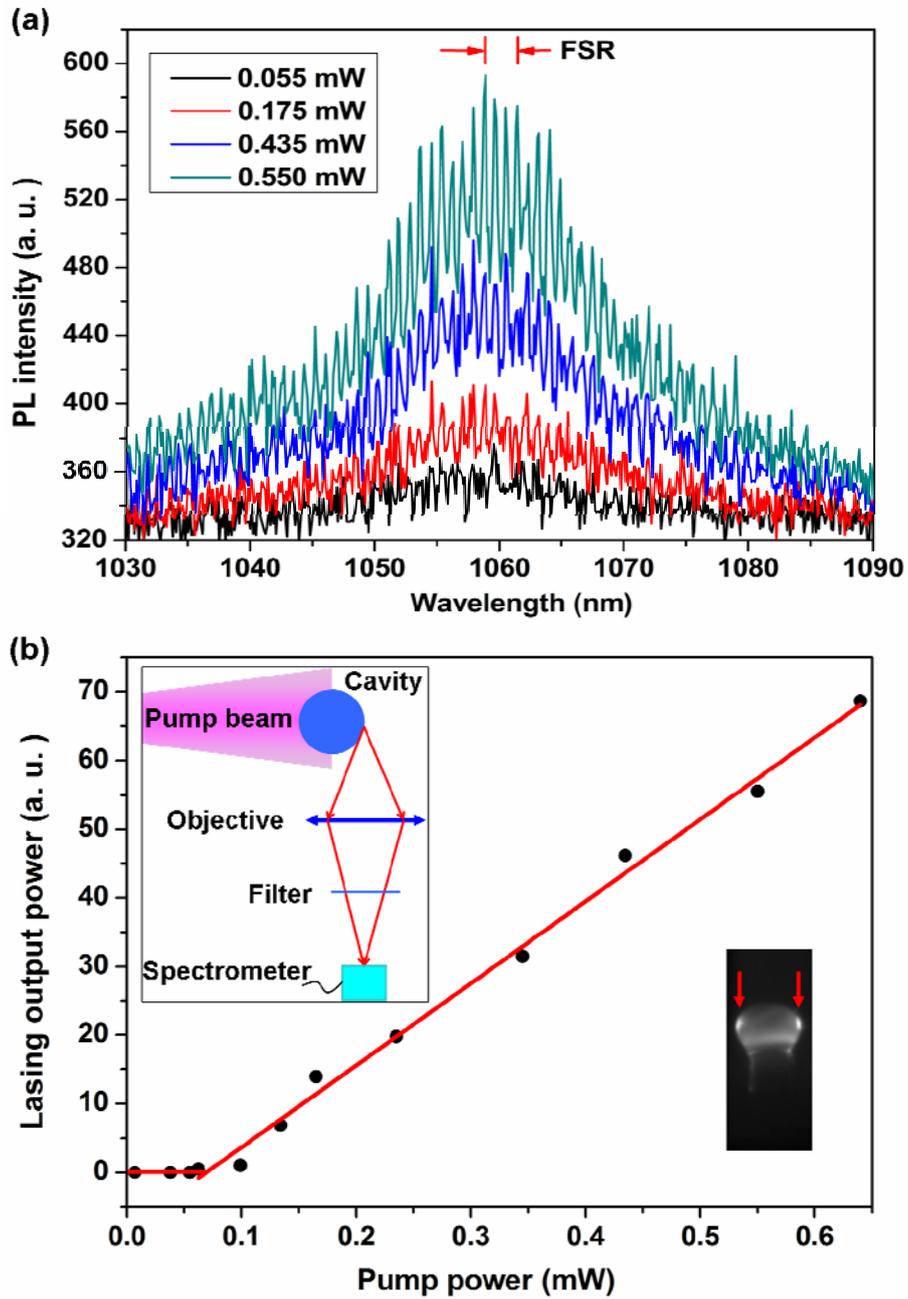